\documentclass[preprint,preprintnumbers,amsmath,amssymb]{revtex4}
\usepackage{graphicx,color}

\begin{document}

\title{Instabilities induced by Light in Liquid Crystal Cells with a Photo-Responsive Substrate}

\author{T. T\'oth-Katona, K. Fodor-Csorba, A. Vajda, I. J\'anossy}

\affiliation{Institute for Solid State Physics and Optics,\\
Wigner Research Centre for Physics,\\
Hungarian Academy of Sciences,\\
H-1525 Budapest, P.O.Box 49, Hungary}

\bigskip

\begin{abstract}
Instabilities are discussed which take place when a nematic liquid crystal (LC) layer, enclosed between a
planar reference plate and a photosensitive substrate, is illuminated with polarized light
from the reference side ({\it reverse} geometry). The dependence of the observed effects on the wavelength,
polarization direction of the light, and on the thickness of the LC layer is explained by a model based on photoinduced
surface torque. The application possibilities of the phenomena are also explored.
\end{abstract}

\maketitle

\vspace*{3mm}

Photoinduced surface reorientation of liquid crystals may occur when one of the plates of the liquid crystal (LC) cell is coated with
a photosensitive material (e.g., with an azo-derivative), and is irradiated by a polarized light \cite{Gibbons1991,Ichimura2000}.
In that case, the director {\bf n} (the unit vector describing the orientational order of the liquid crystal) orients
perpendicular to the light polarization at the photosensitive plate.
Previous studies on this effect have focused on the case when the LC cell is illuminated from the side which is coated with the
photosensitive layer ({\it direct geometry}). In a recent work, however, the {\it reverse geometry} has also been investigated, in which the polarized white light entered from the side opposite to the photosensitive one ({\it reverse geometry}). Both static and dynamic pattern forming instabilities have been observed in this geometry \cite{Janossy2011}.

In the present paper we describe the reorientation process and the instabilities observed under irradiation by monochromatic light.

The photosensitive plate has been prepared by chemisorption of derivatized methyl red (DMR) on the glass surface following the
method developed by Yi {\it et al.} \cite{Yi2009}. For the other (reference) plate of the LC cell commercial, rubbed polyimide-coated slides (E.H.C. Co, Japan) have been used. Cells of thickness of the order of $10 \mu$m have been prepared, and filled with the room temperature nematic 4-cyano-4'-pentylbiphenyl (5CB). In most of the experiments, before and during the filling process the cell has been illuminated with polarized white light (with polarization direction perpendicular to the rubbing direction of the reference plate) from the photosensitive side. This procedure has ensured a good-quality planar alignment as an initial condition. For monochromatic excitation light source green ($\lambda=532$ nm) or blue ($\lambda=457$ nm) lasers have been used. The polarization
direction is described with the angle $\alpha$, where $\alpha=0$ denotes the polarization perpendicular to the rubbing direction
of the reference plate, while $\alpha=90^{\circ}$ is the polarization parallel with it.

\begin{figure}
\centerline{\includegraphics[width=0.5\textwidth]{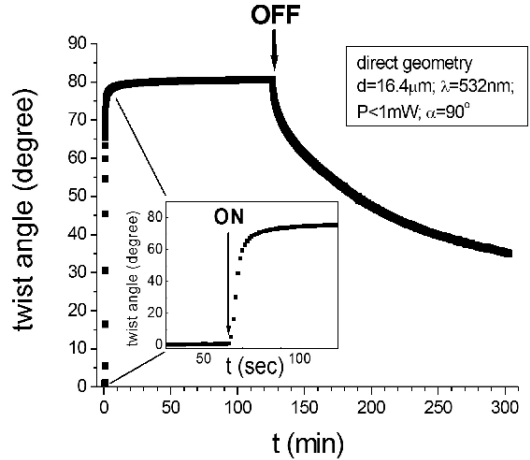}}
\caption{Director reorientation process observed in the direct geometry, with the polarization direction of the
excitation beam $\alpha=90^{\circ}$ which twists the initially planar LC layer. The vertical arrow indicate the time when the excitation beam is switched off, and the relaxation process started. The inset shows enlarged the beginning of the process, when the excitation light is switched on
(indicated again by a vertical arrow).}
\label{fig:Toth_Fig1}
\end{figure}

In the {\it direct geometry} the expected director reorientation process takes place as illustrated in
Fig.\ref{fig:Toth_Fig1}, which shows the temporal evolution of the twist angle for the polarization direction
$\alpha=90^{\circ}$. As one sees, from the initially planar orientation an almost perfect twisted alignment
is achieved within few seconds of illumination. The relaxation process after switching off the excitation beam is, however, much slower.

\begin{figure}
\centerline{\includegraphics[width=0.65\textwidth]{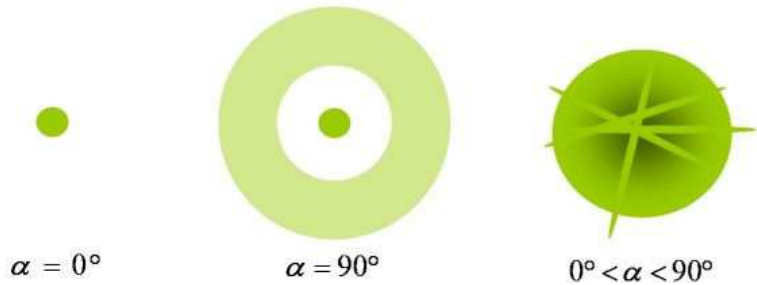}}
\caption{Schematic representation of light scattering patterns observed at different
polarizations of the incoming light in the reverse geometry.}
\label{fig:Toth_Fig2}
\end{figure}

In the {\it reverse geometry} for the planar initial alignment qualitatively different behaviors have been observed depending on the polarization direction $\alpha$ of the excitation beam. The excitation phenomena have been studied by light scattering experiment. The temporal fluctuations of the scattered light intensity have been measured and analyzed recently \cite{Janossy2014}.  The schematic representation of light scattering patterns observed at different
polarizations of the incoming light in the reverse geometry is shown in Fig.\ref{fig:Toth_Fig2}.

In planar cells, the illuminating beam with $\alpha=0$ stabilizes the initial alignment, and therefore, no significant depolarization or scattering has been observed (Fig.\ref{fig:Toth_Fig2}, left panel).

In the case of $\alpha=90^{\circ}$, the scattered light exhibited a broad peak along a cone with an aperture of about
$20^{\circ} - 30^{\circ}$ for $\lambda=532$ nm (Fig.\ref{fig:Toth_Fig2}, middle).
From this aperture a scattering domain size of a few $\mu$m is estimated,
which obviously corresponds to the static domain structure observed previously in a microscope \cite{Janossy2011}.

For the light polarization of $0<\alpha<90^{\circ}$ a dynamic scattering has been observed in the form of random light
intensity both in space and in time. Additionally, flashes of randomly oriented lines also appeared
(Fig.\ref{fig:Toth_Fig2}, right panel).
This dynamic scattering corresponds to the dynamic instability, which has also been observed by a polarizing microscope under
certain conditions \cite{Janossy2011}.

A model has been proposed recently based on photoinduced surface torque, which can account for the above described experimental findings \cite{Janossy2014,Janossy_press}. After making certain assumptions (e.g., monochromatic illumination, light propagation within the Mauguin limit, linearly polarized light on the reference plate at an angle $\alpha$, normal incidence), the photoinduced surface torque $\Gamma_{ph}$ acting at the photosensitive plate becomes

\begin{equation}
\Gamma_{ph} = C I \sin 2\alpha \cos\Delta\Phi
     \label{eq:1}
\end{equation}
with

\begin{equation}
\Delta\Phi = 2 \pi (n_e-n_o)d/\lambda .
     \label{eq:2}
\end{equation}
Here $I$ is the light intensity; $n_e$ and $n_o$ are the extraordinary and ordinary refractive indices, respectively;
$d$ is the sample thickness; $\lambda$ is the wavelength of the light, and $C$ is a constant.

Except the cases which will be discussed below, $\Gamma_{ph}\neq0$ and as a result, the director rotates on the photosensitive plate.
Within the Mauguin limit the polarization ellipse rotates together with the surface director and therefore,
$\Gamma_{ph}$ remains constant in time. On the other hand, the opposing elastic torque increases in time. The balance of torques is, however,
never reached: prior that disclination loops form (as observed in previous studies with polarizing microscopy \cite{Janossy2011}) reducing the twist deformation and decreasing the counteracting elastic torque. We assert that this is the scenario behind the appearance of {\it dynamic instabilities}.

Let us now discuss the cases where $\Gamma_{ph}=0$ according to Eqs.\ref{eq:1} and \ref{eq:2}.

{\bf (i.)} For $\alpha=0$ the surface director at the photosensitive plate is stable against fluctuations of small amplitude, and therefore, the
initial planar alignment remains stable upon illumination.

{\bf (ii.)} For $\alpha=90^{\circ}$ the surface director at the photosensitive plate is unstable against small fluctuations.
Thermally-induced small fluctuations are always present in the system, however, we assert that in this case rather the static orientational disorder on the reference plate is more important \cite{Nespo2010}. Upon illumination the local surface director rotates depending on the local deviation from the average orientation at that particular location of the reference plate. The process leads to formation of static domains that in the same time decrease $\Gamma_{ph}$ and increase the restoring elastic torque. At some stage the balance of torques is reached and the observed {\it static pattern} is formed.

{\bf (iii.)} From Eqs.\ref{eq:1} and \ref{eq:2} follows that the condition $\Gamma_{ph}=0$ can be fulfilled even for polarization directions $0<\alpha<90^{\circ}$ if $\cos\Delta\Phi=0$, i.e., when the thickness of the sample $d$ obeys the relation
 \begin{equation}
 d=(m+\frac{1}{2})\frac{ \lambda}{2(n_e-n_o)} ,
     \label{eq:3}
\end{equation}
where $m$ is an integer. At these thicknesses the director on the photosensitive plate is stable against small fluctuations, and therefore, the dynamic scattering is not expected.

\begin{figure}
\begin{center}
\includegraphics[width=0.65\textwidth]{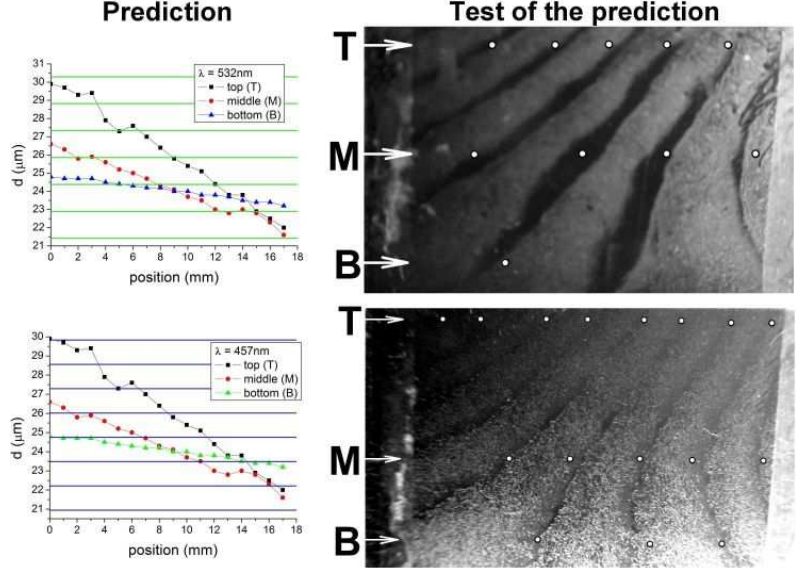}
\caption{Prediction of the model (left) and the results of the test together with the predictions (right). For detailed explanation see the main text.}
\label{fig:Toth_Fig3}
\end{center}
\end{figure}

The prediction {\bf (iii.)} of the model can be tested relatively easily, since $n_e$ and $n_o$ for 5CB, as well as $\lambda$ are known and $d$ can be measured prior filling in the LC. In order to carry out a test, a 2-dimensional wedge LC cell has been prepared in which the thickness varies monotonically both in $x$ and $y$ directions. The thicknesses of the empty cell have been measured by interference method at different positions along $x$ at three different heights $y$ [at the top (T), around the middle (M), and at the bottom (B)]. The results are plotted in Fig.\ref{fig:Toth_Fig3} on the left column. On these graphes the thicknesses that fulfill the condition given by Eq.\ref{eq:3} are also indicated by horizontal lines for both wavelengths of the excitation light $\lambda=532$ nm and $\lambda=457$ nm. After thickness measurements, the cell was filled with 5CB and illuminated by an expanded laser beam with the polarization direction of $\alpha=20^{\circ}$. The resulting patterns are shown in the pictures on the right of Fig.\ref{fig:Toth_Fig3} for both wavelengths, together the predictions of the model where dynamic instability should not occur (indicated by the white spots). The dynamic instability generates disclination loops that scatter the light resulting in opaque regions, while the area without disclinations remains transparent. As it can be seen from  Fig.\ref{fig:Toth_Fig3}, the experimental results support the validity of the model.

\begin{figure}
\centerline{\includegraphics[width=0.65\textwidth]{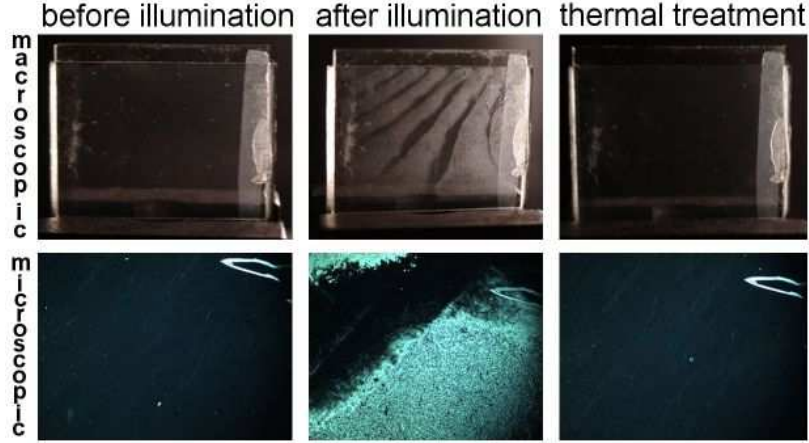}}
\caption{Macroscopic (top row) and microscopic (bottom row) images of the 2-dimensional wedge LC cell before and after the illumination, as well as after the erasing thermal treatment.}
\label{fig:Toth_Fig4}
\end{figure}

It is fascinating that disclination loops generated by the dynamic instability persist for a long time (over weeks). They could be removed completely only by thermal treatment, i.e., by heating the LC cell above the nematic to isotropic phase transition temperature. If the cell is then cooled back to room temperature under simultaneous illumination with a polarized white light having polarization direction perpendicular to the rubbing direction of the reference plate, the original planar orientation is reobtained. This is illustrated in Fig.\ref{fig:Toth_Fig4} with macroscopic and microscopic images of our 2-dimensional wedge cell before and after the illumination, as well as after the thermal treatment. Therefore, in principle, effects described in this paper offer the possibility for a rewritable optical storage device which can be controlled by the light polarization, by the wavelength of the light, or by the sample thickness, and in which the data can be erased by a simple thermal treatment.

In summary, we have shown that light-induced instabilities may occur in the reverse geometry of the LC cell with a photosensitive substrate. A model is presented that explains the observed phenomena, and the potential application possibility of the observed effects is highlighted. As a final remark, we note that a subtle difference between the direct (the light enters the photosensitive layer directly) and reverse geometry (the light goes through the LC layer before entering the photosensitive layer) makes a drastic difference in the observed phenomena.

{\it Acknowledgements.} Financial support by the Hungarian Research Funds OTKA K81250 and NN110672 is gratefully acknowledged.
T.T.-K. is thankful for the hospitality provided in the framework of the HAS-SAS Bilateral Mobility Grant SNK71/2013.


\end{document}